\documentclass[11pt,a4paper]{article}
\pdfoutput=1
\usepackage{jheppub}
\usepackage{amsmath}
\usepackage{verbatim}
\usepackage{amssymb}
\usepackage{inputenc, array}
\usepackage{textcomp}
\usepackage{appendix}

\usepackage{mathrsfs}
\usepackage{cancel}
\usepackage[dvipsnames]{xcolor}
\newcommand{\be}[1]{\begin{equation}\label{#1} }
\newcommand{\ee}{\end{equation}}
\newcommand{\bea}[1]{\begin{eqnarray}\label{#1} }
\newcommand{\eea}{\end{eqnarray}}
\newcommand{\bes}[1]{\begin{subequations}\label{#1} }
\newcommand{\ees}{\end{subequations} }
\newcommand{\p}{\partial}
\newcommand{\refb}[1]{(\ref{#1})}

\title{Stress tensors of 3d Carroll CFTs}

\author[a,b,c]{Sudipta Dutta.}
\author{\\}

\affiliation[a]{Indian Institute of Technology Kanpur, Kanpur 208016, INDIA.\\} 

\affiliation[b]{Institute for Theoretical Physics, TU Wien, Wiedner Hauptstrasse 8–10/136, A-1040 Vienna, Austria\\}

\affiliation[c]{Erwin Schroedinger International Institute for Mathematics and Physics, 1090 Vienna, AUSTRIA}

\emailAdd{dsudipta@iitk.ac.in}

\preprint{}
\abstract{We discuss kinematical features of conformal Carroll field theories in three dimensions (3d). Conformal extension of Carroll algebra is infinite dimensional even in 3d unlike its relativistic counterpart, and hence 3d Carroll CFTs share similarities with 2d CFT. We provide a construction for the conserved charges for Carrollian CFTs and an expression for stress tensor OPEs consistent with the algebra of charges. We discuss a free field model where these symmetries are realised and explicitly compute the OPEs verifying our construction. In addition we comment on the possibility of extending the conformal symmetries to diffeomorphisms of spatial slice for these Carrollian theories.}

\begin{document}
\maketitle

\newpage

%\section*{APPENDICES}

\section{Introduction}

%Gravity is the first fundamental force human beings recognise, yet it remains the least understood one. We understand gravity at the classical level fairly well following Einstein's theory of relativity, but a quantum theory of gravity still eludes physicists. 
%A quantum theory of gravity is essential for understanding the deep questions about the universe's origin and unravelling the mysteries of black holes. 
%Quantum field theory (QFT) provides a natural framework for quantising the other three forces. However, the methods of perturbative QFT are not so successful for gravity due to its bad ultraviolet behaviour. 

Theoretical physicists' best friend in unravelling deep questions about Nature is perhaps the rich group of symmetries physical systems possess. Based on Lorentz symmetries, relativistic quantum field theories successfully describe three of the four fundamental forces. However, the growing number of recent studies shows that several non-Lorentzian symmetries appear in physical systems more often than previously anticipated. One particular symmetry group of interest is the Carroll group. This group was introduced in the sixties by Levy-Leblond and Sengupta \cite{LevyLeblond,NDS} as the ultra-relativistic contraction of the Poincare group. The $c\to 0$ limit indicates a scenario where the light cones close up, forcing the causal particles to confine within a single line. This bizarre wonderland might seem irrelevant at first glance, but many physical systems show this behaviour, ranging from cosmology and gravity to condensed matter systems.

\medskip 

In gravitational physics, the Carroll group is most relevant in the studies of null hypersurfaces. In the intrinsic geometry of these null hypersurfaces, Carroll structures replace the Riemannian ones. Degrees of freedom that live on these null hypersurfaces are naturally Carrollian.
 
\medskip 
 
The most important example of a null hypersurface is perhaps the null boundary of asymptotically flat spacetimes. Quantum field theories on these null hypersurfaces have been proposed as the holographic duals of the gravity theories in flat spacetimes. This approach towards flat space holography \cite{Bagchi:2010zz, Bagchi:2012cy, Bagchi:2016bcd} goes by the name of Carrollian holography in recent literature.  Carrollian holography has indeed met much success in the context of 3-dimensional bulk and two-dimensional boundary theories. This success story principally relies on an isomorphism between the conformal extension of the Carroll group and the BMS group in one higher spacetime dimension. The connection between Carrollian conformal symmetries ($\mathfrak{CCar}$) and BMS symmetries for arbitrary dimensions was clarified in \cite{Duval:2014uva}, following closely related observations in \cite{Bagchi:2010zz}. This isomorphism provides a systematic way to establish a holographic correspondence for flat spacetime in a spirit similar to the usual AdS/CFT correspondence. From the boundary perspective, the zero cosmological limit in the bulk can be seen to be realised as the Carrollian limit. The reader is directed to \cite{Barnich:2006av,Bagchi:2012xr,Barnich:2012aw,Barnich:2012xq,Bagchi:2012yk,Barnich:2012rz,Bagchi:2014iea,Hartong:2015usd,Bagchi:2015wna, Jiang:2017ecm,Hijano:2017eii,Hijano:2019qmi} for a non-exhaustive set of papers in this direction.
 
\medskip 
 
A parallel approach, Celestial holography has emerged from investigation of the infrared sector of flat spacetimes produced many novel results in four dimensions \cite{Strominger:2013jfa,He:2014laa,Cachazo:2014fwa,Kapec:2014opa,Strominger:2014pwa}. It proposes a 2d CFT dual to gravity theories in 4d asymptotically flat spacetimes \cite{Pasterski:2016qvg,Pasterski:2017kqt}. The correlation functions of primary operators of this CFT encode the bulk scattering amplitudes and their infrared properties. Results of celestial holography are excellently summarized in these reviews \cite{Pasterski:2021rjz,Raclariu:2021zjz,Strominger:2017zoo}. 

\medskip 

How a 3d Carrollian field theory can take care of the scattering amplitude was not understood until recently. The readers are directed to \citep{Bagchi:2022emh} for new developments in this direction. A complimentary approach was considered in \citep{Donnay:2022aba}. Further studies on 3d Carrollian CFTs might provide a bridge between these two different approaches towards flat holography.

\medskip

Apart from the holography of flat spacetimes, Carrollian structures also arise in the context of tensionless strings. The string sweeps out a null worldsheet as the tension is dialled to zero. After choosing a suitable gauge, BMS$_3$ appear as a group of residual symmetries, replacing two copies of Virasoro algebra. Some recent advances in classical and quantum tensionless strings, based on these Conformal Carroll or BMS symmetries, can be found in \cite{PhysRevD.16.1722,Isberg:1993av,Bagchi:2013bga,Bagchi:2015nca,Bagchi:2020fpr}.

\medskip

Another important example of a null hypersurface is the event horizon of a black hole. Different versions of BMS symmetries also emerge as near-horizon symmetries \cite{ Grumiller:2019fmp,Donnay:2015abr}. These symmetries and the associated charges have been proposed to account for the entropy of the black holes \cite{Carlip:2019dbu,Carlip:2017xne}. For an explicit realization of Carroll structures on the black hole horizon, the readers are directed to \citep{Donnay:2019jiz,Freidel:2022vjq}. A novel approach towards describing black hole entropy using null strings was taken in \citep{Bagchi:2022iqb}.
  
\medskip  
  
In cosmological scenarios, Carroll symmetries have been found to govern the slow-roll inflationary paradigm \cite{deBoer:2021jej}. Carrollian expansion of gravity has been addressed in \cite{Hansen:2021fxi}. Very recently, Carroll symmetries have also appeared in condensed matter systems, which can be found in \cite{Bagchi:2022eui, Bidussi:2021nmp}. Carrollian fluids have been considered in \cite{Ciambelli:2018wre} with applications to flat holography in mind.

\medskip

Considering the vital role that the Carroll group play in theoretical physics, it is necessary to advance further studies in Carrollian field theories. A selected set of works towards the formal development from intrinsic and limiting perspectives are in \cite{Henneaux:2021yzg,Gupta:2020dtl,Chen:2021xkw,Barnich:2022bni,Bagchi:2019xfx,Saha:2022gjw,Ciambelli:2018ojf}. 

\medskip

In this paper, we consider some generic kinematical features of 3 dimensional conformal Carrollian field theories. Section 2 reviews the essential aspects of the Carroll group and Carrollian manifolds. Section 3 discusses the generic structure of stress tensors for conformal Carrollian field theories and  relation with the BMS charges. We then derive the transformation rules of the stress tensor components under the conformal Carroll group. This essentially provides the same information as the stress tensor OPE. The following section discusses an explicit example of a free scalar for this construction. In section 5, we discuss the possible extension of symmetries for field theory with this structure of stress tensor.

\section{Carroll and Conformal Carroll}\label{car}

\subsection{Carroll group and Carrollian Manifolds}
Carroll group can be obtained by taking $c \to 0$ contraction from Poincare group. This contraction is diametrically opposite to the Galilei case that yields the well known group of Galilean transformation. Like the Galilei group, this opposite contraction also makes perfect group theoretic sense and give rise to the Carroll group. Taking the speed of light to zero would implies a context where the light cones would close up to a single line reflecting the ultralocal behaviour of the theory. 
%This can be explicitly seen from two-point correlation functions of these fields.  
The Carroll group is spanned by usual translation and spatial rotation generators along with the Carroll boosts. 
%For the Carroll invariant theories the boost generator plays a crucial role and makes these theories ultra local.
The Carroll algebra in $(d+1)$ dimensions is given by
\begin{align}
[J_{ab},P_c]&=\delta_{ac}P_b-\delta_{bc}P_a,   \quad [J_{ab},C_c]=\delta_{ac}C_b-\delta_{bc}C_a    \\  \nonumber
[P_a,C_b]&=\delta_{ab}H,   \quad  [C_a,C_b]=0,   \quad [J_{ab},J_{cd}]=\mathbf{•}{so}(d). 
\end{align}
Here $P_a$ and $J_{ab}$ are the translations and spatial rotation generators respectively and  $C_a$ denotes the Carroll boosts.  The Hamiltonian $H$ enters the algebra as a central element only.

\medskip

%\subsection{Carrollian Manifolds}
The kinematical structures associated with this algebra are called Carroll manifolds \cite{Henneaux:1979vn}. They can be obtained by the so called 'gauging' procedure of the Carroll algebra \cite{Hartong:2015xda}. The tangent space structure of these Carrollian manifolds are degenerate and local Lorentz symmetry of Riemannian manifolds are replaced here by local Carroll group. These structures are endowed with a twice symmetric degenerate tensor $h_{\mu\nu}$ and a nowhere vanishing vector field $\tau^\mu$ orthogonal to it. Together they form the notion of metric for Carrollian geometries (i.e invariant objects under tangent space transformations). In a specified coordinate system the Carrollian metric will take the following form
\begin{align}
ds^2=h_{\mu\nu}dx^{\mu}dx^{\nu}=g_{ij}(t,x^k)dx^idx^j  \quad  \tau=\tau^t(t,x^k)\partial_t
\end{align}
where $g_{ij}$  is the non-degenerate $d$-dimensional metric on spatial submanifold and the coordinates $(t,x^i)$ are timelike and spacelike coordinates respectively. 
It is also possible to define the projective inverses  $\tau_{\mu}(t,x^i)$ and $e^{\mu}_a(t,x^i)$  of these metric variables by the following relations
\begin{align}
\tau^\mu\tau_\mu=1, \quad e^{\mu}_ae_{\mu}^a=\delta^{a}_b, \quad \tau_{\mu}e^{\mu}_a=0, \quad e^a_{\mu}\tau^{\mu}=0, \quad \tau_{\mu}\tau^{\nu}+e^a_{\mu}e^{\nu}_a=\delta^{\nu}_{\mu}
\end{align}
Here $(\mu,\nu=0,1,...,d)$ denotes spacetime indices and $(a,b=1,...,d)$ tangent space indices. The degenerate tensor field  can be expressed as $h_{\mu\nu} =e^a_{\mu}e^b_{\nu}\delta_{ab}$. 
%{\tb{What happened to the sentence after this?}}
Under an infinitesimal local Carroll boost and spatial rotation, parametrized by $\lambda^a$ and $\lambda^a_b$, the vielbeins transform like 
\begin{align}
\delta \tau^\mu=0,\quad \delta e^\mu_a=-\tau^\mu\lambda_a+\lambda^a_be^\mu_b, \quad \delta \tau_{\mu}=e^a_{\mu}\lambda_a, \quad \delta e^a_{\mu}=\lambda^a_be^b_{\mu}
\end{align}
These transformation rules can be obtained by assuming the adjoint representation of the local Carroll group in the gauging procedure . Using the above, it is straightforward to verify that  $\tau^\mu(t,x^i)$ and $h_{\mu\nu}(t,x^i)$ remain invariant under the local transformations while their inverses transform in a non-trivial way.

\medskip

%This bizarre wonderland might seem  irrelevant at first glance but they appear quite often in theoretical physics. Especially a class of very useful examples of these Carrollian manifold are null hypersurfaces embedded in Lorentzian space times. Field theories that live on these null hypersurfaces are thus Carroll invariant.

\subsection{Field theory on $\mathscr{I}$}

We are interested in defining quantum field theories that live on the null boundary ($\mathscr{I}$) of asymptotically flat space times.
For the moment, let us consider 4d asymptotically flat spacetimes. Its future null boundary is $\mathscr{I}^+$ , which is topologically $\mathbb{R}_u \times \mathbb{S}^2$ with $\mathbb{R}_u$  as the null direction. We will consider a QFT on this null surface. The degenerate induced metric in this case is given by the line element
\begin{equation}
ds^2=0.du^2+q_{ij}dx^idx^j
\end{equation}
Here $u$ is retarded time coordinate and $x^i$ are coordinates on 2-sphere. The metric on the sphere is denoted by $q_{ij}$. To link up with the Carrollian geometry described in the above section we need to make the following identifications
\begin{align}
\tau^{\mu}=(1,0,0) \quad h_{\mu\nu}=\text{diag}(0,q_{ij}) \quad \text{or} \quad g_{ij}=q_{ij}
\end{align}
The isometry equations on this background are
\be{}
\mathcal{L}_\xi \tau^\mu = 0, \quad \mathcal{L}_\xi g_{\mu\nu} =0, \quad \tau^\mu g_{\mu\nu} =0
\ee
The solutions close to form the Carroll group.  
%\textcolor{red}{(Weak Carroll structures and supertranslations ? )}

\medskip

If we now wish to consider conformal structures, we need to generalise to conformal isometries instead of isometries. The conformal isometry equations on the null surface are given by 
\begin{eqnarray}
	\mathcal{L}_{\xi}g_{\mu\nu}=\lambda(t,x^i) g_{\mu\nu}, \quad \mathcal{L}_{\xi}\tau^\mu=-\frac{\lambda(t,x^i)}{N}\tau^\mu
\end{eqnarray}
The so called dynamical exponent $N$ encapsulates different conformal extensions of the Carroll group, accounting for different relative scaling between space and time. For $N=2$, the space and time dilates homogeneously. This is the case which will be of importance to us throughout the later part of this work. In 3 dimensions, the set of vector fields that solves the above equation for $N=2$ is 
\begin{equation}
\xi=\left[\alpha(x^i)+\frac{u}{2}D_{i} f^i(x^j)\right]\partial_u+f^{i}(x^j)\partial_i \label{kill}
\end{equation}
 where $D_i$ is the connection compatible with $q_{ij}$. $\alpha(x^i)$ is an arbitrary function of $x^i$, but $f^{i}(x^i)$ need to satisfy the following conformal Killing equation on $\mathbb{S}^2$
\begin{eqnarray}
%\partial^i f^j+ \partial^j f^i=\frac{1}{2}q^{ij}\partial_{k}f^k, \quad 
\mathcal{L}_f q_{ij} = D_k f^k \, q_{ij}.
\end{eqnarray}
Choosing the stereographic coordinates $(z, \bar{z})$ on 2 sphere, such that $$ds^2 = \dfrac{2 dz \, d \bar{z}}{(1+z \bar{z})^2},$$ the above equation for the components of $f$ is solved by holomorphic and anti-holomorphic functions, i.e. 
$$ f^z \equiv f^z(z) \quad \text{and} \quad f^{\bar{z}} \equiv f^{\bar{z}}(\bar{z}).$$  Owing to these arbitrary functions the algebra of these vector fields \eqref{kill} is clearly infinite dimensional contrary to its relativistic counterpart and interestingly is also isomorphic to the BMS$_4$ algebra. In a suitably chosen basis the generators of these transformations would look like
\begin{align} \label{kill v}
L_n=z^{n+1}\partial_z+(n+1)z^n\frac{1}{2}u\partial_u, \quad \bar{L}_n=\bar{z}^{n+1}\partial_{\bar{z}}+\frac{1}{2}(n+1)\bar{z}^nu\partial_u, \quad M_{r,s}=z^r\bar{z}^s\partial_u
\end{align}
These generators satisfy the following algebra
\begin{align} \label{Kill al}
[L_n,L_m]&=(n-m)L_{n+m}, \quad [\bar{L}_n,\bar{L}_m]=(n-m)\bar{L}_{n+m}  \\ \nonumber
[L_n,M_{r,s}]&=(\frac{n+1}{2}-r)M_{r+n,s}, \quad [\bar{L}_n,M_{r,s}]=(\frac{n+1}{2}-s)M_{r,n+s}    \\ \nonumber
[M_{r,s},M_{t,u}]&=0
\end{align}
%{\bf{Write the algebra here.}}

\medskip
The relation between the two infinite dimensional symmetry algebras in arbitrary dimension is given by 
\be{}
\mathfrak{CCar}^{N=2}_d = \mathfrak{bms}_{d+1}.
\ee

The killing vectors in \eqref{kill v} are the asymptotic killing vectors of flat spacetimes when projected on to the null boundary. This extension of asymptotic symmetry group was first proposed  in \cite{Barnich:2009se,Barnich:2010eb,Barnich:2011mi}

%\textcolor{red}{Need to decide whether to keep the following:}

\section{Carrollian stress tensors}

The nicest thing about 2d CFT is the infinite extension of the symmetry algebra. The implications of the infinite number of conserved quantitites can be elegantly captured by the holomorphic and anti-holomorphic stress tensor components and their OPEs with the field content of the theory. Although in 3 dimensions this infinite extension of the relativistic  conformal algebra is lost, in the Carrollian regime it's still infinite dimensional. In this section using the stress tensor components we aim to explore how the conformal structures are manifest in 3d Carrollian theories. Consider a field theory coupled to a Carrollian background whose dynamics is governed by an action given by
\begin{align}
\mathcal{S}=\int dt d^2x \, \mathcal{L}(\Phi_i(t,x^i)).
\end{align}
 The stress tensor components can be evaluated by the variation of the action with respect to the vielbeins as
\begin{align}
\delta \mathcal{S}=\int dt d^2x \, e[\tau^{\beta} \delta\tau_{\alpha}+e^{\beta}_a\delta e^a_{\alpha}]T^{\alpha}_{\, \, \beta}
\end{align}

The underlying symmetries of the theory would dictate the form of the stress tensor components. In order to be a conformal Carroll invariant field theory, the covariant action should posses Weyl symmetry along with the invariance under the local Carroll group.

\medskip

\paragraph{Local Carroll invariance:} Under infinitesimal local Carroll group the lower vielbeins transform as 
\begin{equation}
\delta \tau_{\alpha}=e^a_{\alpha}\lambda_a, \quad \delta e^a_{\alpha}=\lambda^a_b e^b_{\alpha}
\end{equation}
using these variations in the above equation we have
\begin{align}
\delta S= \int dtd^2x \, e[(\lambda_a e^a_{\alpha}\tau^{\beta}T^{\alpha}_{\beta})+(\lambda^a_b e^b_{\alpha}e^{\beta}_aT^{\alpha}_{\beta})]=0
\end{align}
On flat Carroll backgrounds this implies 
\begin{align}\label{cart}
T^i_{\, u}=0     \quad   \text{and}    \quad   T^i_j=T^j_i
\end{align}
These are analogous conditions to symmetric stress tensor in relativistic field theories and was previously addressed in \cite{Baiguera:2022lsw}. Although the rotation invariance fixes the spatial part of the stress tensor to be symmetric in $(i,j)$. We shall assume this part to be proportional to the identity matrix, i.e. 
\be{}
T^i_j=k(t,x^i)\delta^i_j.
\ee  
This assumption would be valid for a class of Carrollian theories called 'electric' or the 'timelike' theories where the Lagrangian does not contain any term involving the space derivatives of the field.

\medskip

\paragraph{Weyl invariance:} Under infinitesimal Weyl rescalings the vielbeins would transform as
\begin{align}
\delta \tau^\beta=\lambda(t,x^i) \tau^{\beta}, \quad  \delta e^{\beta}_a=\lambda(t,x^i) e^{\beta}_a
\end{align}

In principle these rescalings might not be homogeneous between timelike and spacelike veilbeins but we shall stick to this case keeping in mind the isomorphism between conformal Carroll and BMS holds for $N=2$, i.e. when the dilatation generator scales space and time in the same way. The variation of action under such rescalings would be
\begin{align}
\delta S=\int dtd^2x \, e\lambda(t,x^i)\left[\tau^{\beta}_{\alpha}+e^{\beta}_ae^a_{\alpha}\right]T^{\alpha}_{\beta}  = \int dtd^2x \, e \lambda(t,x^i)T^{\alpha}_{\alpha}
\end{align}
Thus invariance requires the trace to vanish, i.e $T^{\alpha}_{\alpha}=0$.
Together with the assumptions made previously this tracelessness condition would imply
\be{}
k(t,x^i)=-\frac{1}{2}T^u_u \implies T^i_j=-\frac{1}{2}T^u_u\delta^i_j.
\ee

\subsection{BMS charges}

Here we shall use the standard technique of contracting stress tensors with the killing vectors (\ref{kill}) to evaluate the currents associated with the BMS symmetries
\begin{equation}
J^\alpha=T^{\alpha}_{\beta}\xi^\beta
\end{equation}
Conservation of these currents follow from this specific structure of the stress tensor components and the conformal killing equations
\begin{align}
\nabla_{\alpha}J^{\alpha}=&\nabla_{\alpha}(T^{\alpha}_{\beta}\xi^{\beta}) = T^{\alpha}_{\beta}(\nabla_{\alpha}\xi^{\beta})+\xi^\beta.(\nabla_{\alpha}T^{\alpha}_{\beta})=0 
\end{align}
In the above expression we have used the stress tensor conservation equations and the conformal killing equations.
The corresponding charges can be obtained by integrating $J^u$ over the space slice
\begin{align} \label{blocks}
Q_{\xi}&=\int \sqrt{q}d^2z J^u =\int \sqrt{q}d^2z [T^u_u\xi^u+T^u_i\xi^i]   \\ \nonumber
&=\int \sqrt{q}d^2x^i[T^u_u(\alpha(x^i)+\frac{u}{2}D_if^i(x^j)+T^u_i.f^{i}(x^j)]  \\ \nonumber
&=\int \sqrt{q} d^2x^i [T^u_u.\alpha(x^i)+(T^u_i-\frac{u}{2}D_iT^u_u)f^i(x^j)]
\end{align}

%\underline{a quick look at the stress tensor conservation equations}
To get to the last line of the above equation we have dropped a total derivative term. Now consider the stress tensor conservation equations:
\begin{align}
\partial_uT^u_u+\nabla_iT^i_u=0   \implies \partial_u T^u_u=0. 
\end{align}
In getting to the above, we have used \refb{cart}. We also have:
\begin{align}
  \partial_u (T^u_i-\frac{u}{2}D_iT^u_u) &= \partial_u T^u_i-\frac{1}{2}D_i T^u_u =-D_jT^j_i-\frac{1}{2}D_i T^u_u  =0,
\end{align}
where we use stress tensor conservation and tracelessness. Thus the conservation equations show that both $T^u_u$ and $(T^u_i-\frac{u}{2}D_iT^u_u)$ are independent of $u$. The corresponding charges are hence conserved. From now on, for ease of notation we define 
\be{Tcar}
T_u(z,\bar{z}) \equiv T^u_u, \quad  T_i(z,\bar{z}) \equiv  T^u_i-\frac{u}{2}D_iT^u_u.
\ee 
To get the standard expressions, we decompose the charges by choosing the parameters in this specific way
\begin{align} \label{Charge standard}
L_n=Q_{\xi}[\alpha=0,f^z=z^{n+1},f^{\bar{z}}=0]&=\int d^2z \sqrt{q} T_z. z^{n+1}    \\ \nonumber
\bar{L}_n=Q_{\xi}[\alpha=0,f^z=0,f^{\bar{z}}=\bar{z}^{n+1}]&=\int d^2z \sqrt{q} T_{\bar{z}}. \bar{z}^{n+1}   \\ \nonumber
\text{and} \quad M_{r,s}&=\int d^2z \sqrt{q} T_u.z^r\bar{z}^s
\end{align}

Here $L_n$ and $\bar{L}_n$ are superrotation charges that generate the conformal transformations on $S^2$ and $M_{r,s}$, the supertranslation charges that generate the angle dependent translation along $u$ direction.

\subsection{Charge algebra and stress tensor OPE}
The algebra of these BMS charges would  dictate the transformation rules of these stress tensor components under the BMS group.  The non vanishing commutation relations of BMS$_4$ algebra are given by
\begin{align} \label{Charge al}
[L_n,L_m]&=(n-m)L_{n+m}, \quad [\bar{L}_n,\bar{L}_m]=(n-m)\bar{L}_{n+m}  \\ \nonumber
[L_n,M_{r,s}]&=(\frac{n+1}{2}-r)M_{r+n,s}, \quad [\bar{L}_n,M_{r,s}]=(\frac{n+1}{2}-s)M_{r,n+s}    \\ \nonumber
[M_{r,s},M_{t,u}]&=0
\end{align}
Now using the expressions in (\ref{Charge al}) it can be shown that $T_u(z,\bar{z}),T_z(z,\bar{z})$ and $T_{\bar{z}}(z,\bar{z})$ would transform like  Virasoro primary fields. For simplicity from now on we shall continue with a flat metric of the spatial slice, i.e. $ds^2=dzd\bar{z}$, which would allow us to use $\partial_i$ to $D_i$. Now the left moving Witt algebra would indicate  
\begin{align} 
[L_n, \int d^2z T_z(z,\bar{z}).z^{m+1}]&=(n-m)\int d^2z T_z(z,\bar{z}).z^{m+n+1} \\  
\text{i.e,} \qquad \qquad  [L_n,T_z(z,\bar{z})]&=z^{n+1}\partial_zT_z(z,\bar{z})+2T_z (z,\bar{z})\partial_z(z^{n+1})
\end{align}
 Similarly using the fact $\bar{L}_n$ and $L_m$ commutes, it is possible to derive
 \begin{align}
 [\bar{L}_n,T_z(\bar{z},z)]&=\bar{z}^{n+1}\partial_{\bar{z}}T_z(z,\bar{z})+T_z(z,\bar{z})\partial_{\bar{z}}(\bar{z}^{n+1})  \\ \nonumber
 &=\partial_{\bar{z}}(T_z(z,\bar{z})\bar{z}^{n+1})  \\  \nonumber
 \implies [L_n,\bar{L}_m]&=\int d^2z  \partial_{\bar{z}}(T_z(z,\bar{z}).\bar{z}^{n+1}z^{m+1})=0
 \end{align}
 This transformation rule is different from usual 2d CFT in the sense that the stress tensor components generating the Virasoro subalgebra in this case is not holomorphic anymore. It also has anti-holomorphic weight $\bar{h}=1$ along with $h=2$. Using similar arguments it can be shown that $T_{\bar{z}}(z,\bar{z})$ is also not anti-holomorphic, but transform like a primary with $h=1$ and $\bar{h}=2$.
 
\medskip 
 
 %These are the transformation rules of a primary field with holomorphic and anti-holomorphic weights (2,1) respectively. 

%\tb{Show $T_u$ transformations.}
Using the commutators between $L_n,\bar{L}_n$ and $M_{r,s}$ we can similarly derive the transformation rules of $T_u(z,\bar{z})$. They are
\begin{subequations}
\begin{align}
[L_n,T_u(z,\bar{z})]=z^{n+1}\partial_zT_u(z,\bar{z})+\frac{3}{2}T_u(z,\bar{z})\partial_z(z^{n+1}),   \\ 
[\bar{L}_n,T_u(z,\bar{z})]=\bar{z}^{n+1}\partial_{\bar{z}}T_u(z,\bar{z})+\frac{3}{2}T_u(z,\bar{z})\partial_{\bar{z}}(\bar{z}^{n+1}).
\end{align}
\end{subequations}
Thus from the above analysis we can conclude that the building block of  3d Carrollian stress tensors with conformal symmetry are 3 independent functions of spatial coordinates. The scaling dimension $\Delta=(h+\bar{h})$ can be seen to be equal to 3 as expected for stress tensor components in 3 dimensions. But the underlying algebra dictates the precise way the holomorphic and anti-holomorphic weights split for these objects. This is given in table \refb{t1} below 
 
%Likewise using other components of the algebra it can be proved that $T_u$ and $T_{\bar{z}}$ also transform like Primaries. The weights of these fields are listed below as
\begin{table}[h!]
\centering
\begin{tabular}{||c | c c c||} 
 \hline
 weights &  $T_u$ & $T_z$ & $T_{\bar{z}}$ \\ [0.75ex] 
 \hline\hline
 $h$ &  ${3}/{2}$ & 2 & 1 \\ 
 \hline
 $\bar{h}$ & ${3}/{2}$ & 1 & 2 \\
 \hline
\end{tabular}
\caption{Weights of Carroll stress tensor components}
\label{t1}
\end{table}

Under supertranslations parametrized by $\alpha(z,\bar{z})$ the components of the Carroll stress tensor would transform in the following way
\begin{subequations}
\begin{align}
\delta_{\alpha(z,\bar{z})}T_z=& \frac{1}{2}\alpha(z,\bar{z}) \partial_z T_u(z,\bar{z})+\frac{3}{2}T_u(z,\bar{z})\partial_z\alpha(z,\bar{z}) \\ 
\delta_{\alpha(z,\bar{z})}T_{\bar{z}}=&\frac{1}{2}\alpha(z,\bar{z}) \partial_{\bar{z}} T_u(z,\bar{z})+\frac{3}{2}T_u(z,\bar{z})\partial_{\bar{z}}\alpha(z,\bar{z}) \\
\delta_{\alpha(z,\bar{z})}T_u=&0
\end{align}
\end{subequations}

\medskip

\paragraph{Operator Product Expansions:} In a quantum theory, the transformations of the components of the stress tensor can be equivalently captured in terms of the operator product expansion (OPE) of these components. In  a relativistic 2d CFT, the singular part of the TT OPE is completely fixed by the underlying Virasoro algebra along with some the inputs of an ordering prescription. Owing to the infinite dimensional symmetry algebra the singular part of the stress tensor OPE for three dimensional Carrollian CFTs can also be fixed using the symmetry arguments. Taking hints from the analysis above we propose these stress tensor operator product expansions and then show its equivalence with the algebra. These are 
\begin{subequations}\label{TT}
\begin{align} 
T_z(z,\bar{z})T_z(\omega,\bar{\omega})& \sim 2\partial_{\omega}\delta^2(z-\omega)T_z(\omega,\bar{\omega})+\delta^2(z-\omega)\partial_{\omega}T_z(\omega,\bar{\omega})   \\  
T_z(z,\bar{z})T_u(\omega,\bar{\omega}) &\sim \frac{3}{2}\partial_{\omega}\delta^2(z-\omega)T_u(\omega,\bar{\omega})+\delta^2(z-\omega)\partial_{\omega}T_u(\omega,\bar{\omega}) \\ 
T_z(z,\bar{z})T_{\bar{z}}(\omega,\bar{\omega})& \sim [\partial_{\omega}\delta^2(z-\omega)T_{\bar{z}}(\omega,\bar{\omega})+\partial_{\bar{z}}\delta^2(z-\omega)T_z(z,\bar{z})]   \nonumber
\\  
&+\delta^2(z-\omega)[\partial_{\omega}T_{\bar{z}}(\omega,\bar{\omega})+\partial_{\bar{z}}T_z(z,\bar{z})]  \\   
T_u(z,\bar{z})T_u(\omega,\bar{\omega}) &\sim 0
\end{align}
\end{subequations}

Here $\sim$ denotes only singular terms in the OPEs.  The expressions for $T_{\bar{z}}T_{\bar{z}}$ and $T_{\bar{z}}T_u$  would be similar to the first two expressions of the above equations. These expressions can be shown to be consistent with the charge algebra given in (\ref{Charge al}) using (\ref{Charge standard}). Some of the details of the algebra are given below. The $[L,L]$ commutator can be obtained as: 
\begin{align}
[L_n,L_m]&=\int z^{n+1}d^2z \int  \omega^{m+1} d^2\omega :T_z(z,\bar{z})T_z(\omega,\bar{\omega}):     \\   \nonumber
&=\int d^2z\int (\omega)d^2 \omega [2\partial_{\omega}\delta^2(z-\omega)T_z(\omega,\bar{\omega})+\delta^2(z-\omega)\partial_{\omega}T_z(\omega,\bar{\omega})] \\  \nonumber
&=-\int  d^2z z^{n+m+1}\partial_zT_z(z,\bar{z})-z(m+1)\int  d^2z T_z(z,\bar{z})z^{n+m+1}  \\ \nonumber
&=(n-m)L_{n+m}
\end{align}
The algebra of the super-rotations with the super-translations can be similarly obtained: 
\begin{align}
[L_n,M_{r,s}]&=\int z^{n+1}d^2z \int \omega^r\bar{\omega}^s d^2 \omega T_z(z,\bar{z})T_u(\omega,\bar{\omega}) \\
\nonumber 
&= \int z^{n+1}d^2z \int \omega^r\bar{\omega}^s d^2 \omega[\frac{3}{2}\partial_{\omega}\delta^2(z-\omega)T_u(\omega,\bar{\omega})+\delta^2(z-\omega)\partial_{\omega}T_u(\omega,\bar{\omega})]   \\    \nonumber
&=-\frac{3}{2}r\int d^2z T_u(z,\bar{z})z^{n+r}\bar{z}^s-\frac{1}{2}\int d^2z \partial_zT_u(z,\bar{z})z^{n+r+1}\bar{z}^s  \\  \nonumber
&=(\frac{n+1}{2}-r)M_{n+r,s}
\end{align}
and finally, the two copies of the Witt algebra commute with themselves: 
\begin{align}
[L_n,\bar{L}_m]&=\int z^{n+1}d^2z \int \bar{\omega}^{m+1}d^2\omega T_z(z,\bar{z})T_{\bar{z}}(\omega,\bar{\omega})   \\  \nonumber
&=\int z^{n+1}d^2z \int \bar{\omega}^{m+1}d^2\omega [\partial_{\omega}(T_{\bar{z}}(\omega,\bar{\omega})\delta^2(z-\omega))+\partial_{\bar{z}}(T_z(z,\bar{z})\delta^2(z-\omega))]  \\ \nonumber
&=0
\end{align}
In the above, one noteworthy point is non-vanishing $T_z(z,\bar{z})T_{\bar{z}}(z,\bar{z})$ OPE. This is possible because these stress tensor components carry both non zero holomorphic and anti-holomorphic weights unlike 2d CFT. The presence of the delta functions in the expressions of the OPE can also be traced back to the ultralocal behaviour of the theory. This type of stress tensor OPEs would be expected to appear for the electric theories where the ultralocal behaviour is manifest in the expression of correlation functions. Later on we compute these OPEs by studying an explicit  example of BMS invariant scalar field theory. 

\medskip

We would like to point out that the proposed OPEs \refb{TT} capture only the center-less part of the algebra and we refrain from commenting on the central charge for now. We hope to return to this point in the near future. 

\medskip

\paragraph{Primary fields:} Using this construction the BMS primary fields can be defined in terms of the OPEs of the fields with these stress tensor components. A Carrollian conformal  primary field $\Phi(u,z,\bar{z})$ can be labelled by the eigenvalues of $L_0$ and $\bar{L}_0$ operator, i.e,  
 \begin{align}
 [L_0,\Phi(0)]=h \Phi(0) \quad [\bar{L}_0,\Phi(0)]=\bar{h}\Phi(0)
 \end{align}
and the primary conditions \cite{Bagchi:2016bcd} are
\begin{align}
[L_n,\Phi(0)]=0, \quad [\bar{L}_n,\Phi(0)]=0 \quad \forall n> 0, \quad  [M_{r,s},\Phi(0)]=0  \quad  \forall r,s >0. 
\end{align}
These conditions of primary fields are different from that of a 2d CFT in the sense that half of the supertranslation generators also annihilate  $\Phi(0)$ along with the Virasoro positive modes. This induces the following transformation rules for the primary fields at an arbitrary point at the Carrollian manifold
\begin{subequations}
\begin{align}  \label{Primary}
\delta_{L_n}\Phi(u,z,\bar{z})&=z^{n+1}\partial_z\Phi(u,z,\bar{z})+(h+\frac{u}{2}\partial_u)\Phi(u,z,\bar{z})\partial_z(z^{n+1}),  \\  
\delta_{M_{r,s}}\Phi(u,z,\bar{z})&=z^r\bar{z}^s\partial_u\Phi(u,z,\bar{z}).
\end{align}
\end{subequations}

A similar result holds for $\bar{L}_n$ as well. The tranformation rules under the global generators (i.e. $n=0,\pm 1$ and r,s=0,1) are equivalent to the transformation of momentum space operators of massless particles under the bulk Poincare group. This can be seen by performing a a modified Mellin transformation with respect to the energy variable. The reader is directed to \cite{Banerjee:2018gce,Banerjee:2020kaa} for discussions on modified Mellin representation of the Poincare group. \\
These transformation rules can be encoded in terms of the operator product expansion of the stress tensor components and the operators of concern. Using (\ref{Charge standard}) it can be shown that the above relations are equivalent with
\begin{align}\label{T-Phi}
:T_z(z,\bar{z})\Phi(u,\omega,\bar{\omega}): &\sim (h+\frac{u}{2}\partial_u)\Phi(u,\omega,\bar{\omega})\partial_{\omega}\delta^2(z-\omega)+\delta^{2}(z-\omega)\partial_{\omega}\Phi(u,\omega,\bar{\omega}), \nonumber\\ 
:T_u(z,\bar{z})\Phi(u,\omega,\bar{\omega}): &\sim \partial_u\Phi(u,\omega,\bar{\omega})\delta^2(z-\omega).
\end{align}

\medskip

Having dealt with general consequences, we will in the next section go over to specifics of an example which will provide a robust check of the generic analysis we have laid out in the paper so far.

%\newpage

\section{Free Massless Carroll Scalar Theory}

In the following, we will consider the dynamical perspectives of the Conformal Carroll or equivalently BMS symmetry generators in the context of the free massless Carroll scalar field theory in 2+1 dimensions:
%Let us consider a theory on $\mathcal{I}^+$, given by:
\begin{eqnarray}   \label{Action}
S = \frac{1}{2} \int_{\mathcal{I}^+} du \, d^2 z \,\sqrt{q} \, \dot{\Phi}^2_{h}.
\end{eqnarray}
Here, $\sqrt{q} = \dfrac{1}{(1+z \bar{z})^2}$. This theory is also known as the electric or timelike Carroll scalar. This example of free scalar model was also previously discussed in \cite{Baiguera:2022lsw,Bekaert:2022oeh,Rivera-Betancour:2022lkc,Liu:2022mne}. There is also a magnetic version of the Carroll scalar, which we will not be interested in for the purposes of this work. 

%such that $-2i \dfrac{dz \wedge d\bar{z}}{(1+z \bar{z})^2} = \sin{\theta} d\theta  \wedge d \phi$ is the measure on the celestial sphere, in terms of sterographic coordinates. In the following, we will denote $-2 i dz \wedge d\bar{z}$ as $d^2 z$.% and $d^2 z \equiv i dz \wedge d \bar{z}$.

\medskip

Let us consider the super-rotation generating vector fields (switching off the super-translation):
\begin{eqnarray}\label{suro}
\xi = f^i \partial_i + \dfrac{u}{2}D_i f^i \partial_u
\end{eqnarray}
Keeping in mind that in stereographic coordinates, $f^z$ and $f^{\bar{z}}$ are holomorphic and anti-holomorphic, respectively. %which keeps the degenerate metric on $\mathcal{I}^+$, $ds^2 = \sqrt{q}dz  d \bar{z}$, as well as the null direction $\partial_u$ invariant up to conformal factor.  $D$ is the connection compatible with the Riemannian metric $q$ of $S^2$ and $A$ index runs over $z,\bar{z}$. $Y^z$ and $Y^{\bar{z}}$ are holomorphic and anti-holomorphic functions, independent of $u$.
Infinitesimal conformal diffeomorphism generated by the vector field $\xi$ on field $\Phi (u, z, \bar{z})$ of equal holomorphic and anti-holomorphic conformal weights (since this is a spin-less field) $(h, h)$ is given by:
\begin{eqnarray} \label{HVF}
\delta_{\xi} \Phi_h (u,z, \bar{z}) = f^i \partial_i \Phi + h D_i f^i \, \Phi+ \frac{u}{2}D_i f^i \partial_u \Phi.
\end{eqnarray}

\eqref{HVF} is a Hamiltonian vector field on the space of solutions ($\ddot\Phi = 0$), equipped with the symplectic structure, if $h= \frac{1}{4}$:
\bea{}
\Omega (\delta_1, \delta_2) = \int_{S^2} \sqrt{q} d^2 z \left(\delta_1 \Phi \, \partial_u \delta_2 \Phi - (1 \leftrightarrow 2) \right)
\eea
The corresponding Hamiltonian function is given by
\bea{HF}
&& \Omega( \delta_{\xi},  \delta ) = \delta H[\xi] \nonumber \\
&& H[\xi] = \int_{S^2} \sqrt{q} d^2 z \left(f^i \p_i \Phi\, \dot{\Phi} + \frac{1}{4} u D_i f^i \dot{\Phi}^2 + \frac{1}{4} D_i f^i \Phi \dot{\Phi} \right)
\eea
These are conserved charges as well: $\p_u H[\xi] = 0$. Choosing $f^z = f, f^{\bar{z}} = 0$, the conserved charge reduces to:
\bea{}
Q[f] = \int_{S^2} \sqrt{q} d^2 z \left( f \p \Phi \, \dot{\Phi} + \frac{1}{4 } \left(u\dot{\Phi}^2 + \Phi \dot{\Phi}\right)\frac{1}{\sqrt{q}} \p (\sqrt{q} f )\right)  
\eea
The charge algebra is given by the Poisson bracket:
\begin{equation}
     \{H[\xi_1], H[\xi_2]\} = \Omega (\delta_{\xi_1}, \delta_{\xi_2})   = H[\mathcal{L}_{\xi_1} \xi_2]
\end{equation}
In terms of the holomorphic transformations, we get the centerless Virasoro algebra:
\bea{}
\{Q[f_1], Q[f_2] \} = Q[f_1 \p f_2 -f_2 \p f_1]
\eea
In other words, we define the Virasoro generators:
\bea{}
L_n = Q[f=z^{n+1}]
\eea
to get the standard expression. Using identical methods, we can generate the other copy of the Virasoro in the BMS$_4$ as well. 
Charge corresponding to super-translation $\alpha(z, \bar{z}) \p_u$ is:
\bea{}
Q[\alpha] = \int_{S^2} \sqrt{q} d^2 z \, \alpha (z, \bar{z}) \dot{\Phi}^2
\eea
Choosing $\alpha  = z^r \bar{z}^s$, we get the corresponding $M_{r,s}$ charges, and the BMS algebra is reproduced dynamically at the level of Poisson brackets as expected.

\subsection{Stress tensor}
We will now compute the stress tensor and show its relation with the BMS charges computed above. The stress tensor components computed from (\ref{Action})  following Noether's prescription are
\begin{align}
T^u_u=\frac{1}{2}(\partial_u \Phi)^2, 
\quad 
T^u_i=\partial_{u}\Phi \partial_{i}\Phi, 
\quad
 T^{i}_{j}=-\frac{1}{2}\delta^{i}_{j}(\partial_u \Phi)^2 
\end{align}

The above stress tensor is not traceless. It is possible to improve these components to achieve the prescribed from mentioned in the previous section. The improved set of  stress tensor components are
\begin{align}
T^u_u=\frac{1}{2}(\partial_u \Phi)^2 \quad  T^u_i=\frac{3}{4}\partial_u \Phi \partial_i \Phi-\frac{1}{4}\Phi\partial_u \partial_i \Phi  \quad 
T^i_j=-\frac{1}{4}\delta^{i}_{j}(\partial_u \Phi)^2
\end{align}

This improved stress tensor matches the conditions prescribed in the previous section and thus can be used to compute the BMS charges.This improvement can be accounted for the variation of the conformal coupling term in the covariant action. \cite{Baiguera:2022lsw}

 We can use these improved stress tensor components to evaluate the BMS charges following the prescription mentioned in the previous section as
\begin{align} \label{charge}
%	Q_{\xi}& =\int \sqrt{q} d^2 x^i J^0 \\ \nonumber
%	       &= \int \sqrt{q} d^2 x^i [T_u \xi^u+T^u_i \xi^i]
	      Q_{\xi} &= \int \sqrt{q} d^2 x  \sqrt{q} \left( T_u \xi^u+T_i \xi^i \right) \\ \nonumber
	      &=\int \sqrt{q} d^2x^i \sqrt{q} \Big[ \Big( \frac{1}{2}(\partial_u \Phi)^2(\alpha(x^i)+\frac{u}{2}D_if^i(x^i)\Big) 
	      + \big( (\frac{3}{4}\partial_u \Phi \partial_i \Phi-\frac{1}{4}\Phi \partial_u \partial_i \Phi) f^i(x^i)\Big) \big]
	      \end{align}

%Here we denote $T_u= T^{u}_{u}$ .

We can decompose these expressions by setting $f^i(x^i)=0$ and $\alpha(x^i)=0$ respectively.
\begin{align}
Q[\alpha]&= \int \sqrt{q} d^2 x \sqrt{q} \frac{1}{2} \alpha(x^i) (\partial_u \Phi)^2 
\\ \nonumber
%&=\sum_{r,s} a_{r,s}\int d^2z \sqrt{q} \frac{1}{2}(\partial_u \Phi)^2 z^r \bar{z}^s 
\end{align}
and 
\begin{align}
Q[f]&=\int \sqrt{q} d^2 x  \Big( \frac{u}{4}D_i f^i(x^i)(\partial_u \Phi)^2+ (\frac{3}{4}\partial_u \Phi \partial_i \Phi-\frac{1}{4}\Phi \partial_u \partial_i \Phi)f^{i}(x^i\Big) \\ \nonumber
&=\int \sqrt{q} d^2x  \Big( \frac{u}{4}(\partial_u \Phi)^2. D_i f^i (x^i)-\frac{1}{4}D_i(\Phi\partial_u\Phi)f^i(x^i)
+\partial_u \Phi D_i \Phi f^i(x^i)\Big) \\ \nonumber
&= \int \sqrt{q} d^2x  \Big( \frac{1}{4}
(u(\partial_u \Phi)^2+\Phi \partial_u \Phi\Big) D_i f^{i} + (\partial_u \Phi D_i \Phi ) f^{i}(x^i) \Big)
\end{align}

To arrive at the third line, we have ignored a boundary term. These charges agree with the ones obtained directly using Noether's procedure.

\subsection{Correlators and OPEs}

The Euler-Lagrange equations of motion corresponding to the free Carroll scalar action (\ref{HVF}) is just 
\begin{equation}
\ddot{\Phi}=0
\end{equation}

This equation only allows up to linear term in $u$ in the mode expansions. Generic solution would be 
\begin{equation}  \label{sol}
\Phi(u,z,\bar{z})=A(z,\bar{z})u+B(z,\bar{z})
\end{equation}
Here $A(z,\bar{z})$ and $B(z,\bar{z})$ are arbitrary functions 2-sphere. In terms of these 2d fields $A(z,\bar{z}), B(z,\bar{z})$  the stress tensor components can be re-expressed as
%Transformation rules of $A(z,\bar{z})$ and $B(z,\bar{z})$ under these conformal carroll transformations can be obtained by plugging in the solutions in (\ref{primary}). This yields
%These 2d fields naturally transform as 2d primary operators under the superrotations . This can be verified by plugging in the mode expansions in (\ref{primary}).
%\begin{align}
%\{L_n,A(z,\bar{z})\}=-z^{n+1}\partial_z A(z,\bar{z})-\frac{1}{4}A(z,\bar{z}) \frac{1}{\sqrt{q}}\partial_z(\sqrt{q}z^{n+1})
%\\  \nonumber
%\{L_n,B(z,\bar{z})\}=-z^{n+1}\partial_z B(z,\bar{z})-\frac{1}{4}B(z,\bar{z})\frac{1}{\sqrt{q}}\partial_z(\sqrt{q}z^{n+1})
%\end{align}
%Similar relations hold for the anti-holomorphic case. 
%This shows the solution breaks into two 2 dimensional primary fields with conformal weights $(\frac{3}{4},\frac{3}{4})$ and $(\frac{1}{4},\frac{1}{4})$ respectively.Naturally they can be expanded in z and $\bar{z}$ as
  \begin{align} 
T^u_{u}&=\frac{1}{2}(\partial_u\Phi(u,z,\bar{z}))^2=\frac{1}{2}A(z,\bar{z})^2 \\ \nonumber
T^u_{z}&=\frac{3}{4}\partial_u\Phi(u,z,\bar{z})\partial_z\Phi(u,z,\bar{z})-\frac{1}{4}\Phi(u,z,\bar{z})\partial_u \partial_z \Phi(u,z,\bar{z})  \\ 
&=\frac{u}{2}A(z,\bar{z})\partial_z A(z,\bar{z})+\big(\frac{3}{4}A(z,\bar{z})\partial_z B(z,\bar{z})-\frac{1}{4}B(z,\bar{z})\partial_z A(z,\bar{z})\big)  \\  \nonumber
T^u_{\bar{z}}&=\frac{3}{4}\partial_u\Phi(u,z,\bar{z})\partial_{\bar{z}}\Phi(u,z,\bar{z})-\frac{1}{4}\Phi(u,z,\bar{z})\partial_uD_{\bar{z}} \Phi(u,z,\bar{z})  \\ 
&=\frac{u}{2}A(z,\bar{z})\partial_{\bar{z}} A(z,\bar{z})+\big(\frac{3}{4}A(z,\bar{z})\partial_{\bar{z}} B(z,\bar{z})-\frac{1}{4}B(z,\bar{z})\partial_{\bar{z}} A(z,\bar{z})\big) 
\end{align}

We can identify from the above expressions the '2d'  building blocks of the stress tensors discussed in the previous section. These are
\begin{equation}
T_u=\frac{1}{2}A^2(z,\bar{z}), \quad T_i=\frac{3}{4}A(z,\bar{z})\partial_iB(z,\bar{z})-\frac{1}{4}B(z,\bar{z})\partial_iA(z,\bar{z}),  \quad  i=z,\bar{z}
\end{equation}

\medskip

\paragraph{Correlation functions:} The correlation function of this free scalar theory can be evaluated by computing the Green's functions. The Green's function equation would be
\begin{align}
\partial^2_u G^2(u-u',z^i-z'^i)=\delta^3(u-u',z^i-z'^i)
\end{align}
Solving this equation we find
 \begin{equation} \label{correlator}
 \langle \Phi(u,z,\bar{z})\Phi(u',z',\bar{z}') \rangle=-\frac{1}{2}(u-u')\delta^2(z^i-z'^i)
 \end{equation}
This two-point function is the standard Carroll primary correlator associated with the delta function branch. It is straightforward to check it agrees with the correlator derived from solving the Ward identities \cite{Bagchi:2022emh}, when $\Delta_{\Phi}=\frac{1}{2}$. The delta function appears in the two point function due to the absence of spatial derivatives in the action, and hence in the Green's function equation. This implies transition amplitudes vanish between two spatially separated points. As discussed in the previous sections, this  ultralocal behaviour is quite natural in Carrollian theories. It was also argued in \cite{Bagchi:2022emh} that, the scattering amplitudes are encoded in this branch of Carroll correlators. \\
Using (\ref{sol}) and (\ref{correlator}) we can write 
\begin{align}
 \langle A(z,\bar{z})B(z',\bar{z}') \rangle=\frac{1}{2}\delta^2(z-z',\bar{z}-\bar{z}'), \quad  \langle B(z,\bar{z})A(z',\bar{z}') \rangle=-\frac{1}{2}\delta^2(z-z',\bar{z}-\bar{z}')  
 \end{align}
Below we shall use these correlation functions to compute various operator product expansions by Wick contractions.

\subsection{T-$\Phi$ OPE}
We begin with the $T-\Phi$ OPE. The Wick contractions give us: 
\begin{align}
&:T_z(z,\bar{z})::\Phi(u,z',\bar{z}'):  \nonumber \\ &\sim \frac{3}{4}:A(z,\bar{z})\partial_zB(z,\bar{z})::\Phi(u,z',\bar{z}'):
-\frac{1}{4}:B(z,\bar{z})\partial_zA(z,\bar{z})::\Phi(u,z',\bar{z}'): \nonumber \\ \nonumber
&= \frac{3}{4}[\partial_zB(z,\bar{z})\delta^2(z-z')-uA(z,\bar{z})\partial_z\delta^2(z-z')]  
-\frac{1}{4}[-u\partial_zA(z,\bar{z})\delta^2(z-z')+B(z,\bar{z})\partial_z\delta^2(z-z')] 
\end{align}
So, ultimately we get 
\begin{align}\label{Tp}
T_z(z,\bar{z}) \Phi(u,z',\bar{z}')  \sim \left(\frac{1}{4}+u\partial_u\right)\Phi(u,z',\bar{z}')\partial_{z'}\delta^2(z-z')+\delta^2(z-z')\partial_{z'}\delta^2(z-z')
\end{align}
To arrive at the final expression \refb{Tp}, we have expanded $A(z,\bar{z})$ and $B(z,\bar{z})$ around $(z,\bar{z})$ and also have used the following properties of the delta function: 
\begin{equation}
(z-z')\partial_z\delta^2(z-z')=-\delta(z-z') \quad \partial_z\delta^2(z-z')=-\partial_{z'}\delta^2(z-z').
\end{equation}
The $:T_{\bar{z}}(z,\bar{z})\Phi(u,z',\bar{z}'):$ OPE would be same as $:T_z(z,\bar{z})\Phi(u,z',\bar{z}'):$ with $D_{\bar{z}}$ replacing $D_z$ in the RHS of the final expression \refb{Tp}. Similarly, we can also compute 
\begin{align}
:T_u(z,\bar{z})::\Phi(u,z',\bar{z}'): &\sim \frac{1}{2}:A^2(z,\bar{z})::\Phi(u,z',\bar{z}'): \, = \, :A(z,\bar{z}):\delta^2(z-z') \nonumber
\end{align}
Finally, we get
\begin{align}
T_u(z,\bar{z}) \Phi(u,z',\bar{z}') \sim \partial_u:\Phi(u,z',\bar{z}'):\delta^2(z-z')
\end{align}
These OPEs agrees with (\ref{T-Phi}) derived in the previous section from symmetry arguments with $h,\bar{h}=\frac{1}{4}$.

\medskip

\subsection{T-T OPE}

\medskip

Using these correlators we now compute the operator product expansion of the stress tensor components as well. Here we shall only be interested in the half contracted terms as these terms give rise to the non-central part of the algebra. \\

\medskip
\underline{$:T_z(z,\bar{z})::T_u(z',\bar{z}'):$ }
\begin{align}
:T_z(z,\bar{z})::T_u(z',\bar{z}'):   
\sim & :\frac{3}{4}A(z,\bar{z})\partial_zB(z,\bar{z})-\frac{1}{4}B(z,\bar{z})\partial_zA(z,\bar{z})::\frac{1}{2}:A^2(z',\bar{z}'):   \nonumber \\ \nonumber
=&-\frac{3}{4}:A(z,z)A(z',\bar{z}'):\partial_z\delta^2(z-z') 
+\frac{1}{4}:\partial_zA(z,\bar{z})A(z',\bar{z}')\delta^2(z-z')  \\ \nonumber \\
:T_z(z,\bar{z})::T_u(z',\bar{z}'):  =&\frac{3}{2}T_u(z',z')\partial_{z'}\delta^2(z-z')+\partial_{z'}T_u(z',\bar{z}')\delta^2(z-z')
\end{align} 

\bigskip

\underline{$:T_z(z,\bar{z})::T_z(z',\bar{z}'):$}
\begin{align}
:T_z(z,\bar{z})::T_z(z',\bar{z}'): \sim :(\frac{3}{4}A(z,\bar{z})\partial_zB(z,\bar{z})-\frac{1}{4}B(z,\bar{z})\partial_zA((z,\bar{z})): \\ \nonumber
:(\frac{3}{4}A(z',\bar{z}')\partial_{z'}B(z',\bar{z}')-\frac{1}{4}B(z',\bar{z}')\partial_{z'}A(z',\bar{z}')):
\end{align}
we compute these contractions sequentially below
\begin{itemize}
\item I \begin{align}
&:\frac{3}{4}A(z,\bar{z})\partial_zB(z,\bar{z})::\frac{3}{4}A(z',\bar{z}')\partial_{z'}B(z',\bar{z}'): \\ \nonumber
&=\frac{9}{16}[ \partial_{z'}\delta^2(z-z')(z,\bar{z}')\partial_{z'}B(z,\bar{z}')-\partial_z\delta^2(z-z')A(z,\bar{z})\partial_{z'}B(z',\bar{z}')]  \\ \nonumber
&=\frac{9}{16}[\partial_{z'}\delta^2(z-z'):A(z',\bar{z}')\partial_{z'}B(z',\bar{z}'):+\delta^2(z-z'):A(z',\bar{z}')\partial_{z'}B(z',\bar{z}'):  \\ \nonumber
&-\partial_{z'}\delta^2(z-z'):A(z',\bar{z}')\partial_{z'}B(z',\bar{z}'):-\delta(z-z'):\partial_{z'}A(z',\bar{z}')\partial_{z'}B(z',\bar{z}'):]
\end{align}
\item II
\begin{align}
&:\frac{3}{4}A(z,\bar{z})\partial_zB(z,\bar{z})::-\frac{1}{4}B(z',\bar{z}')\partial_{z'}A(z',\bar{z}'):  \\ \nonumber
=&-\frac{3}{16}[\delta^2(z-z')\partial_zB(z,\bar{z})\partial_{z'}A(z',\bar{z}')-\partial_z\partial_{z'}\delta^2(z-z'):A(z,\bar{z})B(z',\bar{z}'):]   \\ \nonumber
=&-\frac{3}{16}[\delta^2(z-z'):\partial_{z'}A(z',\bar{z}')\partial_{z'}B(z'\bar{z}'):+\partial^2_{z'}\delta^2(z-z'):A(z',\bar{z}')B(z',\bar{z}'):  \\ \nonumber
&+2\delta^2(z-z'):B(z',\bar{z}')\partial_{z'}A(z',\bar{z}')+\frac{1}{2}\partial^2_{z'}A(z',\bar{z}')B(z',\bar{z}'):]
\end{align}
\item III
\begin{align}
&-\frac{1}{4}:B(z,\bar{z})\partial_zA(z,\bar{z})::\frac{3}{4}A(z',\bar{z'})\partial_{z'}B(z',\bar{z'}):   \\   \nonumber
&=-\frac{3}{16}[\delta^2(z-z'):\partial_zA(z,\bar{z})\partial_{z'}B(z',\bar{z'})+\partial_z\partial_{z'}\delta^2(z-z')B(z,\bar{z})A(z',\bar{z'})]  \\ \nonumber
&=-\frac{3}{16}[\delta^2(z-z'):\partial_{z'}A(z',\bar{z'})\partial_{z'}B(z',\bar{z'})-\partial^2_{z'}\delta^2(z-z'):B(z',\bar{z'})A(z',\bar{z'}):
\\ \nonumber
&-2\delta^2(z-z'):A(z',\bar{z'})\partial_{z'}B(z',\bar{z'})-\frac{1}{2}\delta^2(z-z'):A(z',\bar{z'})\partial^2_{z'}B(z',\bar{z'})]
\end{align}
\item IV
\begin{align}
&\frac{1}{4}:B(z,\bar{z})\partial_{z}A(z,\bar{z})::\frac{1}{4}B(z',\bar{z'})\partial_{z'}A(z',\bar{z'}):  \\ \nonumber
=&\frac{1}{16}:[\partial_{z'}\delta^2(z-z'):\partial_zA(z,\bar{z})B(z',\bar{z'})-\partial_z\delta^2(z-z'):B(z,\bar{z})\partial_{z'}A(z',\bar{z'})]   \\ \nonumber
=& \frac{1}{16}[\partial_{z'}\delta^2(z-z'):\partial_{z'}A(z',\bar{z'})B(z',\bar{z'})+\delta^2(z-z')\partial^2_{z'}A(z',\bar{z'})B(z',\bar{z'})]  \\ \nonumber
&-\partial_{z'}\delta^2(z-z'):B(z',\bar{z'})\partial_{z'}A(z',\bar{z'}):-\delta^2(z-z'):\partial_{z'}A(z',\bar{z'}\partial_{z'}B(z',\bar{z'})):]
\end{align}
\end{itemize}
Adding all these contributions together, we get
\begin{align}
& I+II+III+IV   \\ \nonumber
&=\big(\frac{3}{2}A(z',\bar{z}')\partial_{z'}B(z',\bar{z}')-\frac{1}{2}B(z',\bar{z'})\partial_{z'}A(z',\bar{z}')\big)\partial_{z'}\delta^2(z-z') \\ \nonumber
&+\big(\frac{1}{2}\partial_{z'}A(z',\bar{z}')\partial_{z'}B(z',\bar{z}')+(\frac{3}{4}A(z',\bar{z}')\partial_{z'}^{2}B(z',\bar{z}')-\frac{1}{4}B(z',\bar{z})\partial^{2}_{z'}A(z',\bar{z}'))\big)\delta^2(z-z')  \\ \nonumber
\end{align}
So ultimately we have:
\begin{align}
T_z(z,\bar{z}) T_z(z',\bar{z}')  \sim 2T_z(z',\bar{z}')\partial_{z'}\delta^2(z-z')+\partial_{z'}T_z(z',\bar{z}')\delta^2(z-z')
\end{align}
The other Operator product expansions of the stress tensor components can also be shown to hold using the correlation function in a similar fashion. So we have reproduced in a completely different way using Wick contractions in the Carroll scalar the general OPEs which we previous wrote down from symmetry arguments. This provides a robust cross-check of our previous construction. 

\medskip

We remind the reader here that we have not deal with the fully contracted parts of the OPE. These lead to formally divergent terms which when regulated may give us central terms in the corresponding algebra. The regulation scheme is not clear to us at present and this is work in progress and we hope to report on it in the near future.

%\newpage

\section{Enhancement to Diff(S$^2$)?}
In our analysis of the stress tensors of a 3d Carroll CFT, as we showed, under the assumption that the spatial part of the tensor was diagonal, as is expected for electric Carroll theories, there are three components of the stress tensor given by \refb{Tcar}. The charges that we obtained in \refb{Charge standard} were conserved. But a priori, it seems that the expansion \refb{Charge standard} is somewhat ad hoc and restrictive. 

We can define instead define more general charges from the stress tensor components as
\begin{align} \label{Diff charge}
L_{m,n}&=\int \sqrt{q}d^2zT_z(z,\bar{z})z^{m+1}\bar{z}^n , \quad
\bar{L}_{p,q}=\int \sqrt{q}d^2z T_{\bar{z}}(z,\bar{z})z^p\bar{z}^{q+1}.
\end{align}
The algebra of these conserved charges can be checked using (\ref{TT}) and (\ref{Diff charge}). This is 
\begin{align}  \label{Diff $S^2$}
[L_{m,n},L_{a,b}]=(m-a)L_{m+a,n+b}, \quad [\bar{L}_{p,q},\bar{L}_{c,d}]=(p-c)L_{p+c,q+d}  \\ \nonumber
[L_{m,n},M_{r,s}]=\left(\frac{m+1}{2}-r\right)M_{m+r,n+s},  \quad [\bar{L}_{p,q},M_{r,s}]=\left(\frac{q+1}{2}-s\right)M_{p+r,q+s}  \\ \nonumber
[L_{m,n},\bar{L}_{p,q}]=nL_{m+p,n+q}-p\bar{L}_{m+p,n+q} \quad  \text{and} \quad [M_{r,s},M_{i,j}]=0.
\end{align}
The above set of commutators define diffeomorphism extended or generelised  BMS$_4$ algebra \cite{Campiglia:2014yka,Campiglia:2015yka}. The conformal subalgebra, obtained  by setting the appropriate indices to zero, can be identified with the BMS$_4$ algebra given by (\ref{Charge al}). 

\medskip 

Let's try to understand this a bit better. In our previous analysis, we considered only the conformal transformations of celestial sphere at null infinity. It is in principle possible to generalize the symmetries to the whole set of diffeomorphisms of $S^2$ along with the usual supertranslations. The conformal transformations of the celestial sphere are parametrized by the holomorphic and anti-holomorphic functions $f^z(z)$ and $f^{\bar{z}}(\bar{z})$ respectively. So we can relax these holomorphicity conditions and evaluate the charges that generate the diffeomorphisms of $S^2$. These charges given by \refb{Diff charge} would still be conserved in these Carrollian theories if one assumes the structures of the stress tensor prescribed above. This is evident from (\ref{blocks}) as it remains independent of $u$ even if the holomorphicity of the killing vectors are relaxed. The algebra of these charges then closes to the algebra of Diff(S$^2$) \refb{Diff $S^2$}. 

\medskip

At the moment, it seems that the Carroll CFT stress tensors as we have defined them naturally enhance the symmetries from the superrotations to Diff(S$^2$) giving us another larger set of conserved charges. So the pertinent question to ask is which of the two infinite dimensional symmetry enhancements should we go with. At first glance, the superrotation enhancement seems to be the more natural choice since this is the conformal symmetries of the background and it is not clear what we mean when we consider the full diffeomorphism algebra on the sphere when we are considering CFTs on null backgrounds. This expectation is further borne out by a rather straightforward analysis. 

\medskip

Following our previous analysis of the OPEs of stress tensors and Carroll primaries, we can now consider the action of the Diff(S$^2$) generators on the Carroll primary fields. We consider the $T(z,\bar{z})\Phi(u,\omega,\bar{\omega})$ OPE in (\ref{T-Phi}) to concretely write down the transformation rules of the fields by the action of the diffeomorphism charges: 
\begin{align}
\delta_{L_{m,n}}\Phi(u,\omega,\bar{\omega})&=\int \sqrt{q}d^2z T_z(z,\bar{z})\Phi(u,\omega,\bar{\omega}) z^{m+1}\bar{z}^n \\  \nonumber
&=(h+\frac{u}{2}\partial_u)\Phi(u,\omega,\bar{\omega})D_{\omega}(\omega^{m+1})\bar{\omega}^n+D_{\omega}\Phi(u,\omega,\bar{\omega})\omega^{m+1}\bar{\omega}^n.   \\  \nonumber
\delta_{\bar{L}_{p,q}}\Phi(u,\omega,\bar{\omega})&=(\bar{h}+\frac{u}{2}\partial_u)\Phi(u,\omega,\bar{\omega})D_{\bar{\omega}}(\bar{\omega}^{q+1})\omega^p+D_{\bar{\omega}}\Phi(u,\omega,\bar{\omega})\bar{\omega}^{q+1}\omega^p.
\end{align}
The transformation rules of $\Phi(u,\omega,\bar{\omega})$ under supertranslations remain as usual. However, as pointed out in \cite{Schwarz:2022dqf}, these primary transformation rules are not consistent with the symmetry algebra in (\ref{Diff $S^2$}) as the following Jacobi identity does not close for primary fields with non-zero spin.
\begin{align}
&[[L_{p,q},\bar{L}_{r,s}],\Phi(u,\omega,\bar{\omega})]+[[\bar{L}_{r,s},\Phi(u,z,\bar{z})],L_{p,q}]+[[\Phi(u,z,\bar{z}),L_{p,q}],\bar{L}_{r,s}] \\ \nonumber
&=qr(h-\bar{h})\Phi(u,z,\bar{z})z^{k+m}\bar{z}^{l+n}
\end{align}
There is no problem with the closure of Jacobi identity if we choose not to extend the conformal symmetry to diffeomorphisms by setting $q,r=0$. 

\medskip

So, at least with the primary transformation rules prescribed as above, it seems there are problems in extending the super-rotations to the whole of Diff(S$^2$) in the context of 3d Carrollian CFTs. 

\medskip

In \cite{Schwarz:2022dqf}, it is further shown that one fix this issue by the addition of a spin operator $S_{k,l}$ of weight $(h, \bar{h}) = (1,1)$ in the algebra which repairs the Jacobi identity. It is unclear how one would be able to construct such an operator in a 3d Carrollian theory. But the implications of this enhancement to Diff(S$^2$) on scattering are rather severe and calls for a conservation of helicity and is hence unnatural. However this enhancement of symmertries in case of Carrollian scalar theories could still be useful in other purposes. We don't immediately have anything further to add to this.

\section{Discussions}

\subsection{Summary of results}
In this paper, we have studied kinematical features of conformal Carroll invariant field theories in three dimensions. We discussed that local Carroll and Weyl symmetry forces the stress tensor to take a certain form. Using this specific form of the stress tensor, we then constructed the conserved charges associated with the BMS or conformal Carroll transformations. The algebra of the charges dictates the transformation properties of these stress tensor components. We found that three independent stress tensor components transform as primary fields with holomorphic and anti-holomorphic weights $(\frac{3}{2},\frac{3}{2}),(2,1),(1,2)$ respectively. The equivalent statement was then made in terms of stress tensor operator product expansions. Consequently the singular part of the OPEs are also shown to be consistent with the BMS$_4$ algebra. The set of stress tensor OPEs, consist of contact terms and hence reflect the ultralocal nature of Carroll CFTs. This is the main result of this paper. Also it is worthwhile to mention that  although the left and right moving Virasoro generators commute, the associated generating stress tensor components have a non trivial operator product expansion, contrary to 2d CFTs. Furthermore we evaluate what the Carrollian primary conditions translate to in terms of T-$\Phi$ OPEs. In the next section, we discuss an example of free scalar model where this construction was explicitly realised, providing a robust cross-check of the more general analysis earlier. We evaluated the conserved charges for this theory and showed their relations with the stress tensors. More importantly  we also computed the stress tensor OPEs directly by the method of Wick contractions, using the novel ultralocal branch of Carroll correlators. These OPEs also agree with previous construction. In the last section we discussed the possibility of extending conformal symmetry to diffeomorphisms in the context of Carrollian framework and its potential difficulties.

\subsection{Connections to flat holography}

One of the main motivations behind our construction is obviously applications of these 3d Carroll CFTs to gravitational physics in asymptotically flat spacetimes through the holographic principle. We have not directly connected with this in the main body of our paper here. There are various points to clarify here and comment on another potential application of this framework. 

\paragraph{Charge conservation:} The BMS (and Diff(S$^2$)) charges we have defined in this work are conserved. They don't have any dependence on $u$. Now if we are thinking of spacetimes where there is radiation leaking out of $\mathscr{I}^+$ and we envision a field theory capturing this, it is not immediately obvious if conservation of charges are well equipped for that purpose. 
%Instead of this, we can postulate that a Carrollian CFT, not necessarily living on the null boundary but which takes into account two different Carroll CFTs on $\mathscr{I}^\pm$ , captures the whole of bulk flat space physics. 
However unitarity is clearly present in the bulk and would hence manifest itself on any holographic dual field theory. Whatever radiation leaks out of $\mathscr{I}^+$ must have had come from incoming particles through $\mathscr{I}^-$. In short, what goes in must come out. It is plausible that even in asymptotically flat spacetimes with radiation, a dual theory based on Carrollian CFTs would capture the bulk physics and have conserved charges. Let us remember that Carroll CFTs do encode bulk scattering in a non-trivial way as well. Of course, in this case, the explicit way that in-flux and out-flux behaves, or equivalently the information of the Bondi news tensor is lost. 

\paragraph{Soft factorisations:} Scattering amplitudes in the bulk is the  observables of primary interest for flat space holography. The infrared sector of scattering matrices in flat spacetime is quite rich and subtle and elegantly described by the infrared triangle. The infinite number of conserved BMS charges causes the scattering amplitudes to factorise into lower point amplitudes . This factorisation of amplitudes is famously known as the soft factorisation. In order to be a holographic dual of gravity theories in asymptotically flat spacetimes, the boundary field theory must be able to somehow capture the scattering amplitudes and its infrared properties. Recently it has been proved that the these Carrollian field theories can indeed describe the bulk scattering matrix. Having this framework at hand, it would be of immediate interest to investigate the infrared sector from field theory side. The conserved BMS currents constructed in this work and associated Ward identities can be expected to be relevant in capturing bulk infrared physics. It would also be of interest to relate this framework to the conformally soft sector of celestial CFTs. 

\section*{Acknowledgements}
 We are indebted to Arjun Bagchi for his valuable insights at different stages of this work and comments on the draft. We would also like thank Shamik Banerjee, Rudranil Basu, Daniel Grumiller, Romain Ruzziconi and Amartya Saha for various fruitful discussions. SD is supported by grant number 09/092(0971)/2017-EMR-I from Council of Scientific and Industrial Research (CSIR), India and the Junior Research Fellowship Programme from ESI Vienna. SD would also like to acknowledge the hospitality of Institute for Theoretical Physics, TU Wien and Erwin Schroedinger International Institute for Mathematics and Physics, Vienna, where a significant part of this work was carried out. These results were discussed in an online seminar at Ecole Polytechnique, Paris. we would like to thank Andrea Puhm and the other participants for useful discussions following the talk.

\bibliographystyle{JHEP}
\bibliography{ref.bib}
\end{document}